%% file: hind-pc6.tex
\documentclass[aps,prb,superscriptaddress,reprint]{revtex4-1}
\usepackage{graphicx}
\usepackage{latexsym}
\usepackage{amsmath}
\usepackage{amssymb}
\usepackage[permil]{overpic}
\usepackage[colorlinks]{hyperref}
\hypersetup{allcolors=blue}
\begin{document}
\title{Dielectric constant of disordered phases of the smallest monoalcohols : evidence for the hindered plastic crystal phase.}
\author{M.V.Kondrin}
\email{mkondrin@hppi.troitsk.ru}
\affiliation{Institute for High Pressure Physics RAS, 108840 Troitsk, Moscow, Russia}
\author{A.A. Pronin}
\affiliation{General Physics Institute RAS, 117942 Moscow, Russia}
\author{Y.B. Lebed}
\affiliation{Institute for Nuclear Research RAS, 117312 Moscow, Russia}
\author{V.V. Brazhkin}
\affiliation{Institute for High Pressure Physics RAS, 108840 Troitsk, Moscow, Russia}
\begin{abstract}
With gradual temperature increase  in premelting regions of solid phase of methanol and high pressure phase of ethanol, and using novel procedure of separation of electrode polarization effects, we are able to register the contribution of relaxation process to low-frequency dielectric constant. This contribution is about half  the liquid's dielectric constant  near temperature  of solidification , and is almost an order of magnitude higher than reported earlier for ambient pressure phase of methanol. As opposed to  dielectric constant of water at ambient pressure, which does not change much  during crystallization, our finding indicates the hindrance of molecule rotation in orientationally disordered phases of monoalcohols. Similar dielectric responses of ambient pressure methanol and high-pressure phase of  ethanol imply  existence of hindered plastic crystal phase of ethanol (not observed at low pressures). We also have found some dynamic disorder in nominally fully ordered phase of these monoalcohols ($\alpha$-phase of methanol and low-pressure phase of ethanol), the contribution of this disorder being dependent on external conditions (e.g. temperature), and increasing at approaching  the order-disorder transition. On the other hand, the amplitude of dielectric responce in hindered plastic crystal phases is almost independent of temperature.
\end{abstract}
\maketitle

\section{Introduction}
The concept of plastic crystal was first introduced by Timmermans almost hundred years ago \cite{timmermans:jpcs61}. It was based on observation, that some molecular crystals with almost globular molecular structure have melting entropy lower than 5 cal/ K mole (20.92 J/K mole). This implies that  some thermodynamic degrees of freedom (predominantly rotational ones) are already activated in solid phase, thus leading to reduced value of fusion entropy. It was postulated, that  high-temperature solid phases of these substances just below the melting curve consist of orientationally disordered molecules, which  can freely rotate around translationally ordered positions.  Translationally ordered, but orientationally disordered crystal phase was named plastic crystal. Thermodynamic criterion of plastic crystals was chosen slightly arbitrary, thus, for example, ordinary ice with fusion entropy of  21.97 J/ K mole, with  hydrogen bonds disorder, does not belong to them. However,   thermodynamic notion clearly discerns between molecular substances in which orientational disorder at increase of temperature precedes  the onset of translational one (melting).

\begin{figure}
\begin{overpic}[width=\columnwidth]{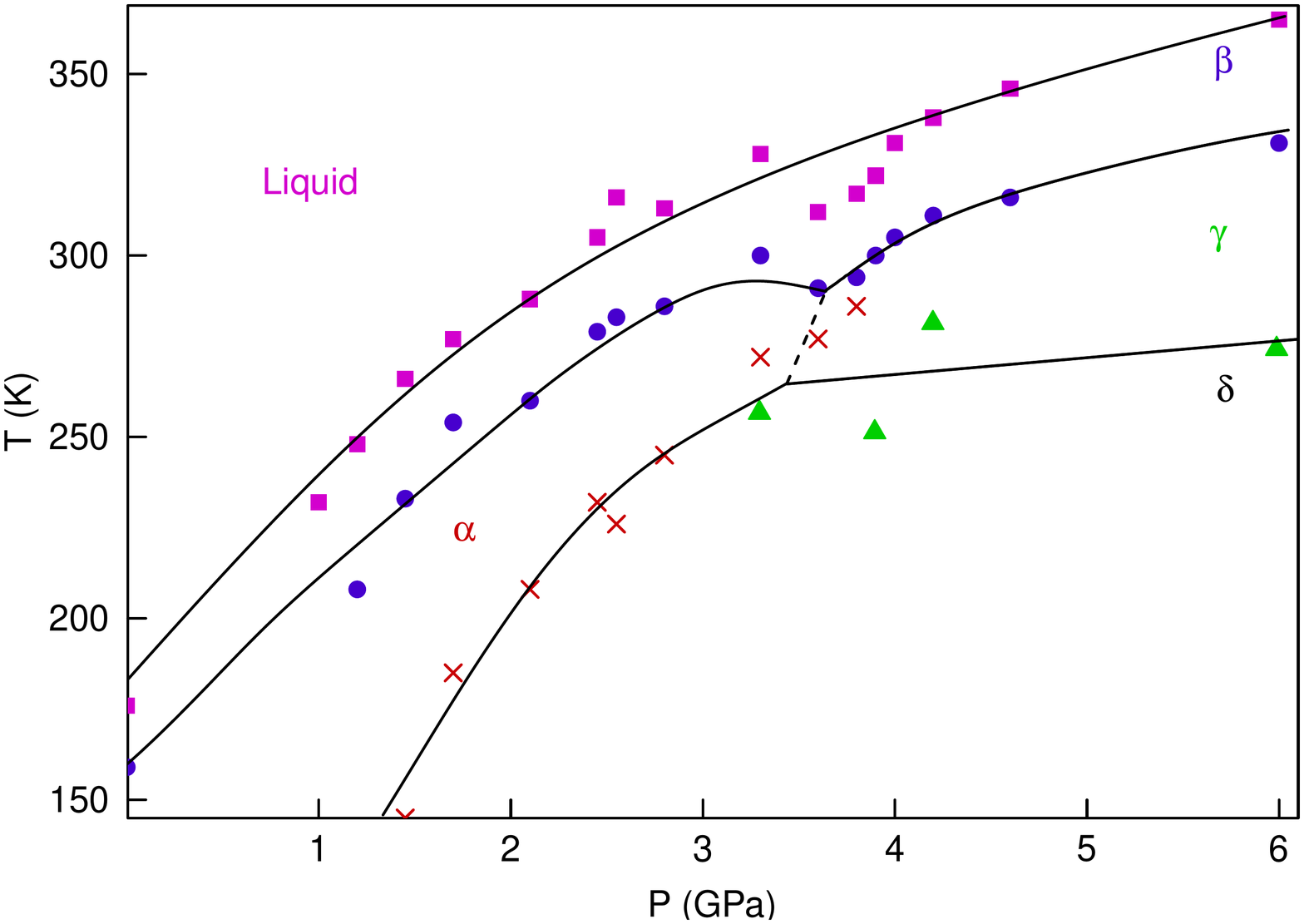}
\put(100,730){{\Large a)}}
\put(650,150){\includegraphics[width=0.25\columnwidth]{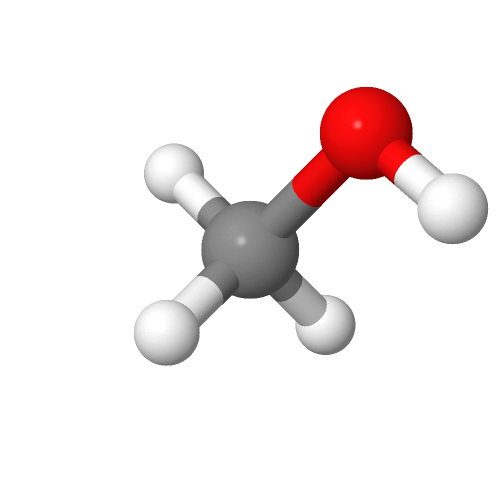}}
\end{overpic}
\begin{overpic}[width=\columnwidth]{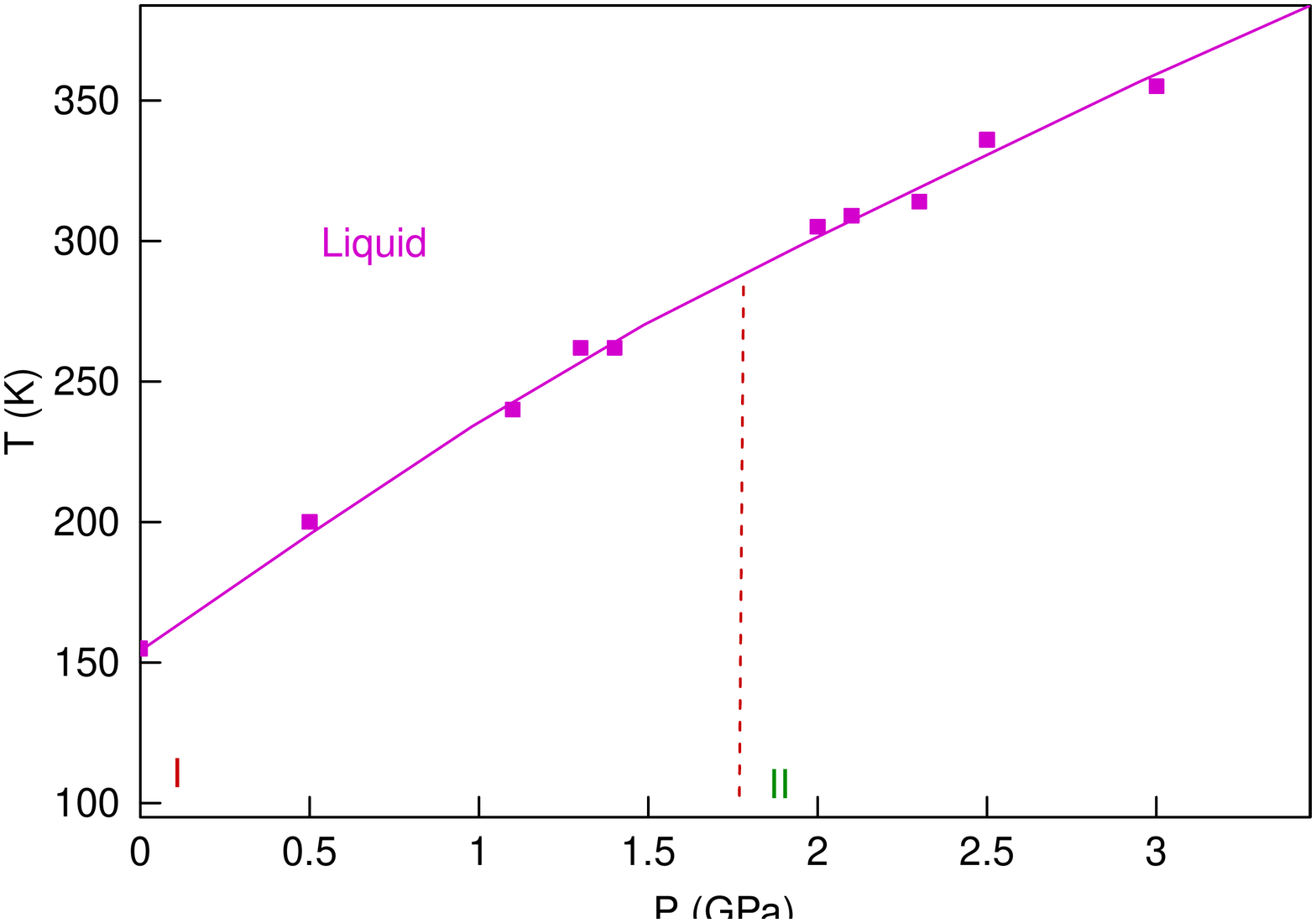}
\put(100,730){{\Large b)}}
\put(650,150){\includegraphics[width=0.25\columnwidth]{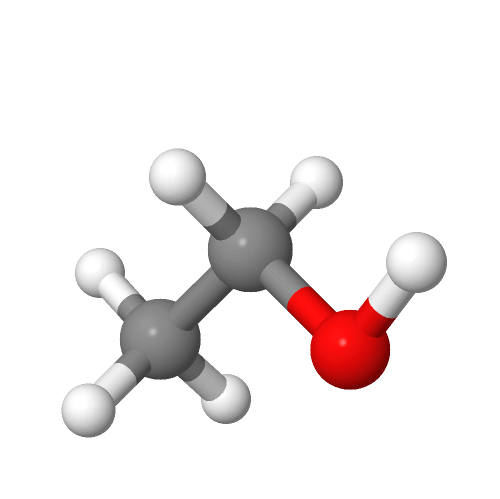}}
\end{overpic}
\caption{a) Phase diagram of methanol according to Kondrin {\em et al.} \cite{kondrin:jcp13}. $\beta$-phase corresponds to orientationally disordered methanol phase. b) Phase diagram of ethanol according to Kondrin {\em et al.} \cite{kondrin:jcp14}. In insets the respective molecules of monoalcohols are shown.}
\label{f0}
\end{figure}

 Crystallographic investigations  are more  definite and can discern  between various types of  disorder. While orientational disorder in hexagonal ice persists from 273 K down to helium temperature,  orientationally disordered methanol ($\beta$-phase) exists in quite limited temperature range -- from 175.5 K to 160 K. Moreover, the model predicts  different types of disorder in ice and methanol. In ice, hydrogen bonds can switch between four positions in  tetrahedrally bonded hexagonal crystal \cite{onsager:jcp69,ryzhkin:jetpl20}, while, in $\beta$-phase of methanol, hydrogen bonds have linear structure (plastic crystal of methanol has anisotropic orthorhombic structure with $Cmcm$ space group \cite{tauer:ac52,allan:prb98,torrie:mp89,torrie:jssc02} in contrast to highly symmetric hexagonal ice) and can be switched between  two positions only. It implies different degrees of hindrance of molecules' free rotation  in the phase of plastic crystal. Crystallographic studies reveal  disorder of hydrogen atoms in  $\beta$-phase of solid methanol, while C-O axis direction being fixed \cite{tauer:ac52,allan:prb98,torrie:mp89,torrie:jssc02}. The remnant of this linear ordering in monoalcohols persists even in the liquid phase which results in difference between characteristic time of dielectric response and structural relaxation \cite{gainaru:prl10,bohmer:pr14}. In ordinary ice, on the contrary, only oxygen position is fixed, and the pair of protons can freely select between four bonds with nearby oxygen atoms (so called Bjerrum fault) \cite{onsager:jcp69,ryzhkin:jetpl20}. Previous dielectric spectroscopy studies have shown that disordered phase of methanol exists till very high pressures ($P <$ 6 GPa, see Fig.~\ref{f0} a)  \cite{kondrin:jcp13}, while in ice the pressure increase  results in exceptionally rich phase diagram which includes crystallographically different proton ordered and disordered phases. In case of methanol theoretical studies (classical molecular dynamics \cite{gonzalez:jcp16} and {\em ab-initio} crystal structure prediction \cite{cervinka:cs18}) are of little help, because although they describe more or less precisely the stability region of ordered methanol phases ($\alpha$ and $\gamma$), but both unanimously predict metastability of orientationally disordered $\beta$-phase ,  which contradicts experimental observations.

Since hydrogen bond has large dipole moment, the difference in crystal structures (and accompanying disorder) would greatly influence the value of dielectric constant \cite{onsager:jcp69}. Static dielectric constant is sensitive to fluctuations of molecular dipole moment, so with the switches of molecular dipoles  its  value would be large, while in frozen molecules only induced polarization ( an order of magnitude weaker) can contribute to dielectric response (thus leading to low values of dielectric permittivity). In liquid water at 273 K dielectric constant is equal to 81 and slightly increases in solid phase \cite{ellison:jpcrd07,johari:jcp81}. This implies that almost free rotation of molecular dipoles is practically retained in solid phase. On the other hand, dielectric constant of liquid methanol phase is equal to $\approx$ 75 near the melting point, but strongly decreases  during crystallization (down to 4-6) \cite{denney:jcp55,davidson:cjc57,kondrin:jcp13}. This implies strong hindrance of free rotation in plastic crystal phase of solid methanol, which is not compatible with crystallographic evidence of almost free rotation of methanol molecules around C-O bond \cite{tauer:ac52,allan:prb98,torrie:mp89,torrie:jssc02}. We argue, that static dielectric constant of methanol in the disordered $\beta$ phase is significantly larger, but corresponds to the low frequency relaxation process, so it almost merges with effects caused by contact polarization processes. Still, in premelting region, we are able to separate these two types of processes. Comparison of obtained dielectric spectroscopy data for methanol with  high pressure data obtained for another small-molecule monoalcohol, ethanol, enables us to suggest the existence of orientationally disordered crystal phase in ethanol at high pressures.

\section{Methods}

Dielectric spectroscopy measurements of ethanol and methanol at ultra-high pressures  were carried out in Toroid chamber \cite{hpr:khvostantsev2004}according to routine, reported earlier \cite{pronin:pre10,pronin:jetpl10,kondrin:jcp12}. Experimentally accessible frequency window (10 Hz -- 2 MHz) and precision of dielectric susceptibility measurements ($\Delta \varepsilon \approx 0.1$) were determined using QuadTech-7600\cite{quadtech:7600}device .  The values of capacitance were measured twice before an experiment , in empty and filled capacitor, and subsequently used for calculation of dielectric susceptibility at high pressure. ``Empty'' capacitance readings were about 10 pF.  During  high-pressure experiments, variation of thermodynamic parameters in crystal phase was possible only along isobars (with typical rate $\pm 0.5$ K/min), because otherwise the pressure change would lead to the breakage of the measurement capacitors. The pressure values during experiments were estimated by the readings of manganin gauge in liquid phase , just prior to the crystallization onset, with typical accuracy of $\pm$ 0.05 GPa. The rate of temperature variation was not strictly controlled, mostly  it was determined by the thermal inertia (quite large) of our setup. Typical cooling rate was about  $-0.5$ K/min, but  heating rate could vary  in the range $0.5 - 1$ K/min by application of external manual heating. The temperature of the measurement  cell was controlled by alumel-chromel thermocouple with precision of 0.01 K. The results presented  were obtained in  several  high-pressure experiments with different measurement cells.  Ambient pressure measurement was carried out using controlled heating/cooling setup and was effectuated with the rate $\pm$ 0.01 K/s. The data eported  were collected during heating of previously cooled samples.

Complex dielectric response of a substance at frequency $\nu$ can be obtained from specific capacitance $C(\nu)$ and conductivity $\sigma(\nu)$  using the formula:

\begin{equation}
\varepsilon(\nu)=\varepsilon'-i \varepsilon''=\frac{C(\nu)}{\varepsilon_0}+ \frac{\sigma(\nu)}{i 2 \pi \varepsilon_0 \nu}
\end{equation}

Physical contribution to this response at low frequencies consists of several inputs, including  static ionic conductivity and intrinsic dielectric permittivity  of the sample with characteristic relaxation frequency $\nu_0$ and additional constant  dielectric permittivity at ``infinite'' frequency. They are usually described by the respective terms:

\begin{equation}
\varepsilon(\nu)=i\frac{\sigma_0}{\nu}+\Delta \varepsilon\mathcal{R}(\nu/\nu_0) +\varepsilon_\infty
\label{eps}
\end{equation}

The contact polarization  term \cite{ishai:mst13} is usually modeled with constant phase element with impedance:
\begin{equation}
   Z_p \propto (i \nu)^{-n}
\label{eq:zp}
\end{equation}
  
(where $0< n<1$) connected in series with the sample. However, this approximation should not be taken too seriously, because beside strong deviation of dielectric permittivity from constant at low frequency region (which is  indeed observed in practice), it also results in unphysical decrease of capacitance in high frequency region. So, one has to provide for a certain cutoff function, able to describe faster fall off of dielectric responce at higher frequencies (which is yet to be accomplished). Moreover, this approach is usually applied to liquid samples, and it is not clear if it can describe polarization of the sample in the solid form. This significantly undermines separation of the terms, related to contact polarization, from intrinsic sample relaxation and makes impossible direct fitting of experimental data with  the  formula proposed. However, since intrinsic relaxation can be described by purely Debye function in monoalcohols at low pressures  \cite{bohmer:pr14}:

\begin{equation}
\mathcal{R}(\nu/\nu_0) = \frac{1}{\left(1 + i\nu / \nu_0 \right)}
\label{cd1}
\end{equation}

it is possible to determine inflexion point on the dependence:

\begin{equation}
\frac{1}{Z 2 \pi \nu}=\frac{1}{2 \pi \nu (Z_p+1/(\varepsilon 2 \pi \nu))}=\frac{\sigma_{eff}(\nu)}{2 \pi \nu}+i \varepsilon_{eff}(\nu)
\label{eq:znu}
\end{equation}

which means that the response of sample and electrode polarization terms, connected in series, is approximated by effective leaky capacitor with frequency dependent conductivity $\sigma_{eff}$ and dielectric function $\varepsilon_{eff}$ . This inflexion point marks transition from diverging region, induced by electrode polarization effects at very low frequencies, to constant term at somewhat higher frequencies, caused by the sample dielectric  relaxation and conductivity. Typical dielectric response, obtained by Equation (\ref{eq:znu}) is depicted in Fig.~\ref{f0}. 

\begin{figure}
\includegraphics[width=\columnwidth]{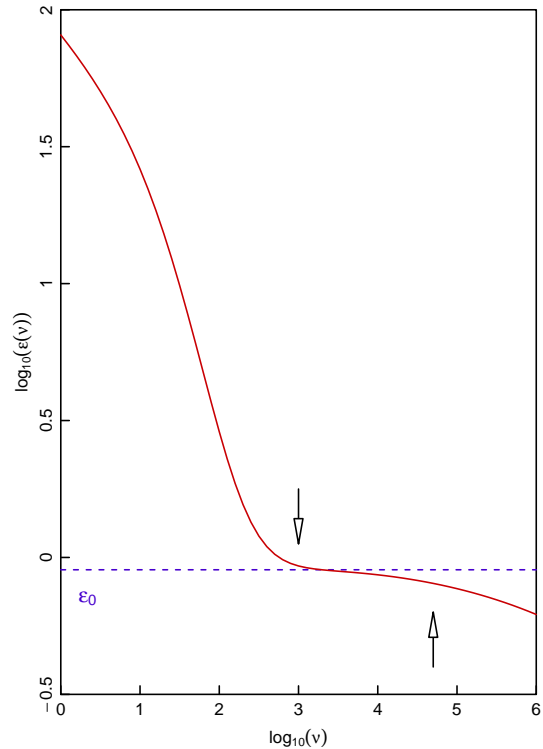}
\caption{Effective dielectric function, caused by polarization effects. Downward and upward arrows mark inflexion points due to the onset of deviations, caused by electrode polarization effects in low frequency and high-frequency region respectively. The sample was modeled, as element with constant conductivity and capacitance (equal to 1). Polarization impedance  was approximated as constant phase element with exponent 0.7. Dashed horizontal line marks position of ``apparent'' static dielectric constant of the sample (it is slightly less than the true one).}
\label{f0}
\end{figure}

Although it is very interesting to separate the frequency input of bulk solid sample from polarization effects, it is well beyond the scope of present paper. Still, gradual application of temperature with simultaneous collection of dielectric permittivity data, applied in our experiments, helped us to successfully separate these two inputs in premelting temperatures range. This topic will be thoroughly discussed later on in Section \ref{meoh:ep}. However, we should stress once more, that our ``continuous heating'' approach is very different from previous works on methanol and ethanol, where frequency scans at fixed temperature were used.

\section{Dielectric spectroscopy of methanol}

\begin{figure}
\includegraphics[width=\columnwidth]{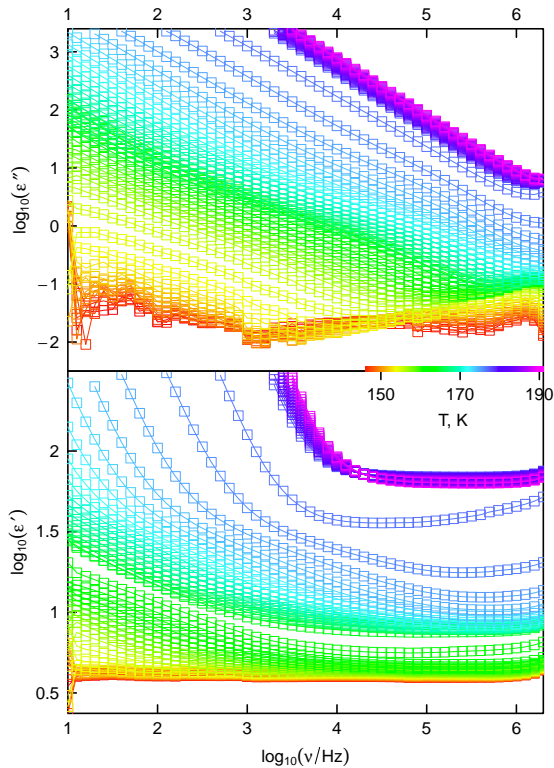}

\caption{Dielectric response of methanol at ambient pressure at specified temperatures.}
\label{f1}

\end{figure}

\begin{figure}
\includegraphics[width=\columnwidth]{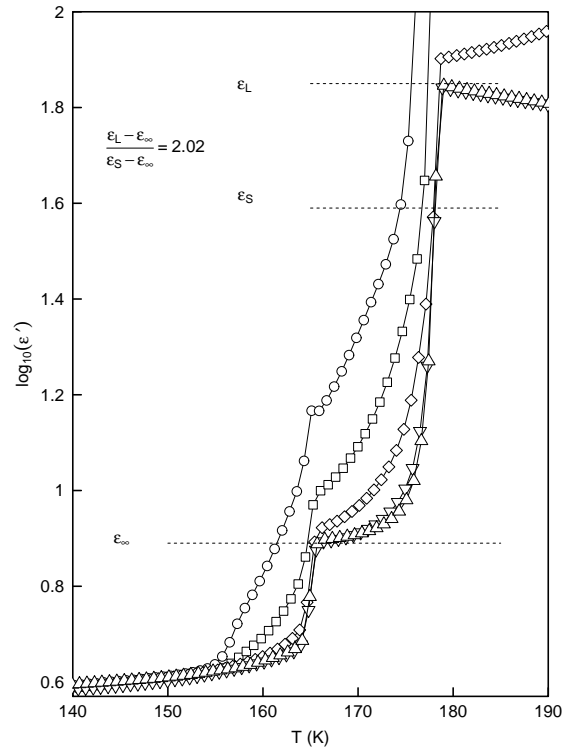}
\caption{Temperature dependence of ambient pressure dielectric permittivity of methanol at various frequencies ($\circ$--100 Hz, $\square$-- 1 KHz, $\lozenge$-- 10 kHz, $\triangledown$ -- 100 kHz,  $\triangle$-- 1 MHz). Note the  log scale of $y$-axis. }
\label{f2}
\end{figure}

Dielectric relaxation of methanol at ambient pressure at scans with constant rate along temperature and frequency is presented in Fig.~\ref{f1} (the scans were collected at heating of previously crystallized sample). Another presentation of the same data, as slices at fixed frequencies, is shown in Fig.~\ref{f2}. The main conclusion that can be drawn from these data is, that freezing of methanol (at 180 K) leads to significant drop of dielectric constant ($\varepsilon '$). Still, in solid phase of methanol, dielectric response is frequency dependent, so the notion of dielectric constant in this case needs some clarification. Large increase of dielectric constant at frequencies below 1 kHz is due to contact polarization and can be regarded as parasitic effect. This effect is also temperature dependent, and the onset of contact polarization frequency is increased to 10 kHz near the melting temperature. Still the dispersion observed in the frequency range 10 kHz-2 MHz can be considered as intrinsic and connected to relaxation of the solid, orientationally disordered methanol ($\beta$-phase). The presence of disordered beta phase in methanol persists till very high pressures (at least up to 6 GPa \cite{kondrin:jcp13}). 

Frequency slices at 100kHz, 1 MHz (see Fig.~\ref{f2}) distinctly demonstrate transition at 165 K from orientationally disordered $\beta$-phase to orientationally ordered low-temperature $\alpha$-phase of methanol, which is accompanied by the drop of high frequency $\varepsilon '$ from 7.7 to 3.9. This is in $\approx$ 1.5 margin corresponds to the results of Denney and Cole \cite{denney:jcp55} ( they report dielectric response in the disordered phase of methanol in the range 4-6 at 5 K below the melting temperature and fast increase of dielectric permittivity at approaching the melting temperature). Nonetheless, in the intermediate temperature region corresponding to the $\beta$-phase, dispersion of dielectric response is observed, which can not be attributed to contact polarization. It manifests itself as distinct step on the frequency dependent response of $\varepsilon '$ (see Fig.~\ref{f1}). This step shifts to lower frequencies with the decrease of temperature   (Fig.~\ref{f1}). Extrapolating this frequency dependent response to low frequency region, we would get static dielectric permittivity, approximately equal to $\varepsilon_S = 38.9$. Further lowering of temperature results in transition to  fully ordered phase, where amplitude of low-frequency relaxation process drops to zero. It should be noted that the similar behavior in methanol below the melting temperature was observed in the work of Denney and Cole \cite{denney:jcp55}. They carried out experiments in setup with significantly larger separation of electrodes ($\approx$ 1 mm vs. 0.1 mm in our case) so in their setup influence of electrode polarization effects should be lower and restricted to lower frequencies. Still, they report observation of dielectric dispersion in the range below 10$^5$ Hz which they attributed to the intrinsic dielectric response of the sample. They also report significant deviation of the dielectric response depending of the thermal history, which can be due to the anisotropic character of the solid phase (which the authors of Ref.~\cite{denney:jcp55} seems to support) but as well to electrode polarization in the solid phase.

Near the melting temperature the amplitude of this relaxation rapidly increases (up to values $\varepsilon_L \approx 70.8$) which is also accompanied by rapid increase of frequency of corresponding relaxation process. Thus, relaxation process rapidly moves out of the experimental frequency window and dielectric response of the liquid methanol is characterized by constant, but very high dielectric permittivity caused by fast free fluctuation of molecular dipoles \cite{denney:jcp55,davidson:cjc57,bertolini:jcp83}. It is interesting to note, that the amplitude of dielectric process is increased by the value approximately equal to 2 $(\varepsilon_L-\varepsilon_\infty)/(\varepsilon_S-\varepsilon_\infty)=2.02$. We argue, that this increase is connected to increased hindrance on the molecules' free rotation in liquid and the solid $\beta$-phase of methanol.

\begin{figure}
\includegraphics[width=\columnwidth]{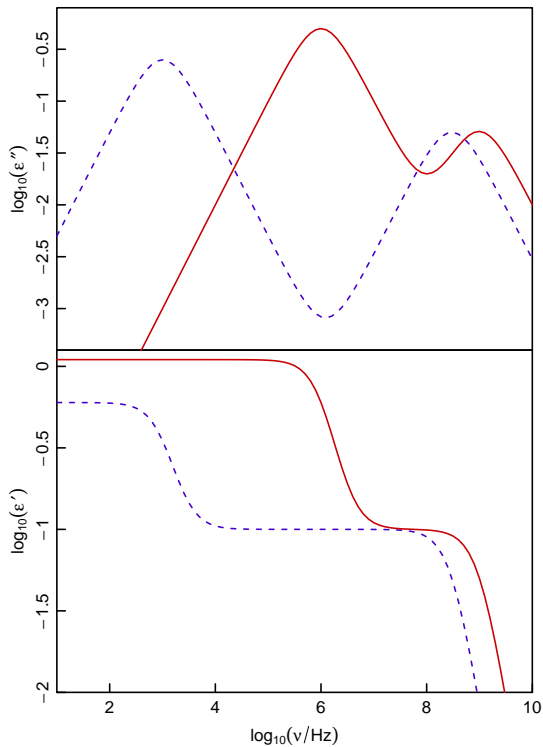}
\caption{Dielectric relaxation in liquid (solid curves) and  disordered solid phase (dashed curves) of the model substance, having two types of relaxation (low and high frequency ones), which represents  dielectric response of methanol in the liquid and plastic crystal solid phase.}
\label{f3}
\end{figure}

In the following we will need some model of dielectric relaxation in monoalcohols. It is known to consist of two processes --the larger effect observed at low frequencies is  Debye relaxation ( at ambient pressure relaxation can be well approximated by the Debye relaxation law) and the weaker, high frequency one , called structural or $\alpha$-relaxation. It is interesting to note, that in more complex (in comparison to methanol) monoalcohols, the frequency of this  weak relaxation usually corresponds to processes, which govern thermodynamic and structural properties observed at {\em e.g.} vitrification of monoalcohols \cite{bohmer:pr14,danilov:jpcb17} . Thus, despite its weakness in thermodynamic response, it is more important in description of molecular dynamics of monoalcohols. Although in application to methanol, exact P-T evolution of the Debye- and $\alpha$-relaxations is not well known (which is mostly due  to the fact, that methanol vitrifies only at high pressures  \cite{brugmans:jcp95,kondrin:jcp13} where precise spectroscopic techniques are not possible), but our findings suggest that,  separation into these  two process exists in methanol too. It can be concluded also from the experimental data  of Barthel {\em et al.} and Fukusawa {\em et al.} \cite{barthel:cpl90,fukusawa:prl05} where at room temperature the separation between the Debye relaxation process and the next in frequency (presumably $alpha$ one) reaches almost one decade (3.09 Ghz and 22.4 Ghz respectively). 

To make this model more tractable, we propose simplified picture, describing dielectric relaxation in liquid and solid phases of methanol Fig.~\ref{f3}. Note, that dielectric losse $\varepsilon ''$ connected to dielectric permittivity $\varepsilon '$ through the Kramers-Kroning relations so the absorption peaks in the experimentally inaccessible high-frequency regions manifests itself in the low-frequency region as plateaus in the dielectric permittivity values. The picture shown in Fig.~\ref{f3} suggests, that the frequency separation between Debye and structural relaxation increases during solidification of methanol (the frequency of the Debye process significantly drops in the solid phase), while amplitude of the Debye relaxation decreases in comparison to  liquid state. This is in contrast to water, where amplitude of dielectric relaxation remains constant in the melting process, while frequency of Debye relaxation is dropped by several orders of magnitude \cite{johari:jcp81,ellison:jpcrd07,aragones:jpca11}. We consider this difference between water and methanol as a manifestation of different types of disorder, present in orientationally disordered solid phases. While in solid  water (ordinary hexagonal ice), the molecule can jump between four positions and so the dielectric constant has the same value as in liquid phase, where molecular dipole can freely rotate, in $\beta$-phase of methanol (on the other hand) molecular dipoles can take only two collinear positions, so corresponding dielectric permittivity value decreases in comparison to liquid state. The rise of orientational disorder during transformation of $\beta$-methanol into fully orientationally ordered phase would lead to suppression of both relaxations (low- and high frequency ones) and subsequently to  drop of dielectric permittivity to lower values, which is indeed  observed in our experiments. 

This difference is obviously connected to presence of methyl group, serving as an anchor, preventing from molecule rotation in solid phase. However, substitution of methyl by ethyl group ( as in ethanol) leads to drastically different behavior in the low and high pressures which will be discussed later in Sections~\ref{et:lp},\ref{et:hp}.  

\section{Separation of electrode polarization effects from the bulk dielectric response in methanol at ambient pressure}
\label{meoh:ep}

We would like to diverge on the topic of separation of electrode polarization effects from the response of the sample in the case of liquid and solid methanol. As it follows from Eqs.~(\ref{eps}-\ref{cd1}) the overall impedance of the sample is the sum of two inputs: the ``bare'' sample impedance and the impedance due to the electrode polarization which can be only roughly approximated by constant phase element. Here, we will demonstrate that the separation of these two effects in the series of experiments obtained at the one sample under varying external conditions ({\em e.g.} varying temperature) does not require laborious multiparametric fitting and can be carried out by the determination of position and the value of only one critical point.

As it follows from Equations~(\ref{eps},\ref{eq:zp},\ref{cd1}) in the low-frequency region ( where the input of this two contributions is approximately equal) the relaxation part and constant current conductivity can be described as leaky capacitor with frequency independent dielectric constant $\varepsilon$ and conductivity $\sigma$. The impedance of such element is given by the formula:

\begin{equation}
Z(\nu)=\frac{1/\varepsilon}{\sigma/\varepsilon + i 2 \pi \nu}= \frac{A}{B+i \omega}
\end{equation}

I
maginary part of this formula has distinct peak-like shape. The peak maximum is located at $\omega=B$ and its maximum reaches the value $\frac{A}{2B}$. It turns out (as it will be demonstrated below) that the value of this maximum (in the vicinity of the maximum) is significantly higher than the impedance caused by electrode polarization effects given by the Eq.~(\ref{eq:zp}) by almost order of magnitude. So, determination of the value and position of this maximum would allow us to determine the static conductivity and the low-frequency dielectric constant of the sample itself.

\begin{figure}
\includegraphics[width=\columnwidth]{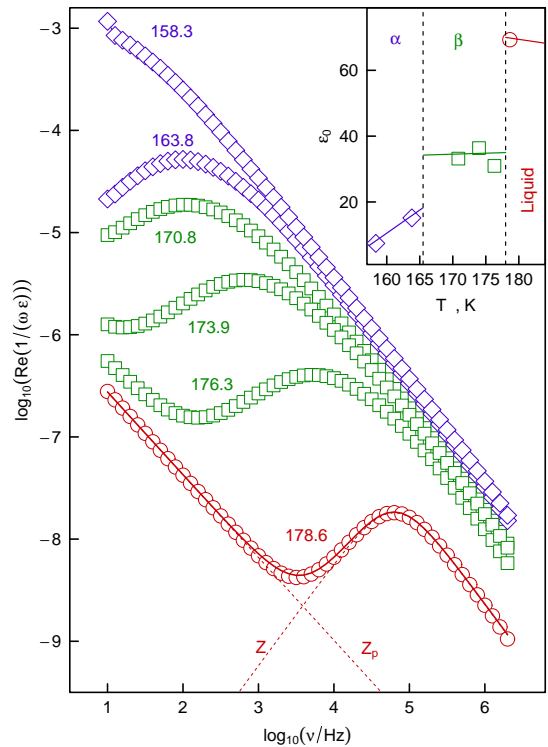}
\caption{Frequency dispersion of the real part of experimental impedance $1/(\varepsilon \omega)$ of methanol at ambient pressure near the melting temperature and $\alpha-\beta$ phase transition. $\circ$-- corresponds to the dielectric response in the liquid phase, $\square$-- in the solid $\beta$ phase, $\lozenge$ -- in the solid $\alpha$ phase. The temperatures in K at which these data were collected is shown near the respective dependencies. Dashed thin curves demonstrate deconvolution of dielectric response in the liquid phase of methanol into electrode polarization part ($Z_p$) and the part due to the bulk of the sample $Z$. Solid thick line is the contribution from both parts $Z_p+Z$. In the inset the temperature dependence of static dielectric constant determined from the local maxima of frequency dependence of impedance is shown. The symbols shape corresponds to the respective curves in the main panel. The vertical dashed lines marks temperatures of phase transitions in methanol determined in our experiments. Solid lines are just guides to the eyes and the temperature trends they depict may be exaggerated.   }
\label{f3a}
\end{figure}

The experimental dielectric response of methanol converted into impedance values at several fixed temperatures in vicinity of the melting region and $\alpha-\beta$ phase transition is depicted in the main panel of Fig.~\ref{f3a}. Aforementioned maximum on frequency dependency of impedance is evident in all these curves. Moreover, in the high temperature limit (in the liquid state and just below liquid-$\beta$ phase transition) it is possible to separate the input to overall dielectric response from the sample itself and the electrode polarization effects. In the liquid state this enables us to fully deconvolve the dielectric response into these two inputs. This yields the exponent $n$ of electrode polarization of methanol in the liquid state which has the value $n=0.81$. Variation of temperature influences both electrode polarization and the sample dielectric response but from position of local maximum we can extract data on the temperature dependence of static dielectric constant near the melting temperature. This data is depicted in inset in Fig.~\ref{f3a}.

First of all, from this data it follows that the ``naive'' separation of constant $\varepsilon$ pertinent to the sample's bulk by simply ignoring divergent tail pertinent to the electrode polarization effects really works under condition that the dielectric constant of the sample is large enough (like in the liquid and the solid disordered $\beta$-phase of methanol). Both approaches yield approximately equal values of dielectric constant: $\varepsilon \approx 70$ for liquid state and constant (independent of temperature) $\varepsilon \approx 35$ in the disordered $\beta$-phase. However, the application of this procedure to the ordered $\alpha$-phase produce rather unexpected result. It turns out that the dielectric constant at $\alpha-\beta$ transition drops to rather large value $\approx 15$ but further diminishes at the temperature lowering (and reaches the value of $\approx 7$). This suggests that the dielectric constant in the fully ordered phase of methanol increases when the temperature approaches the order-disorder phase transition in the solid state. So, this means that some amount of dynamic disorder is present even in the nominally fully ordered phase in the vicinity of phase transition and temperature rising increase the amount of this disorder. The similar behavior is observed in the low-pressure solid phase of ethanol in vicinity of melting temperature which will be discussed in the next Section~\ref{et:lp}. 

\section{Dielectric spectroscopy of ethanol at low pressures}
\label{et:lp}
Ambient pressure solid phase of ethanol (marked as I in Fig.~\ref{f0} b) can by no means  be considered as orientationally disordered. Beside crystallographic evidence \cite{jonsson:ac76,allan:prb99,allan:jsr01}, it  does not by large margin satisfy Timmerman's criterion (its fusion entropy is equal to 31 J/K mole). Similar conclusions can be derived from dielectric spectroscopy experiments at ambient pressure, where transition from $\varepsilon'$ values equal to 75 at the melting temperature to 4-5 in the fully solidified samples \cite{hassion:jcp55,brand:prb00,kondrin:jcp14}. The last value of dielectric permittivity is compatible with the notion of induced polarization of ethanol molecules with fully arrested molecular motion.

\begin{figure}
\includegraphics[width=\columnwidth]{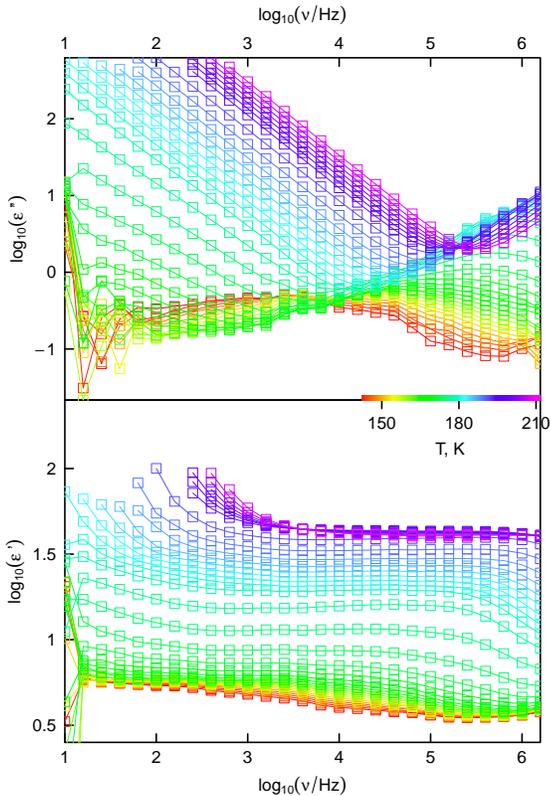}
\caption{Experimental dielectric relaxation in liquid and solid low-pressure phases of ethanol at 0.5 GPa.}
\label{f4}
\end{figure}

\begin{figure}
\includegraphics[width=\columnwidth]{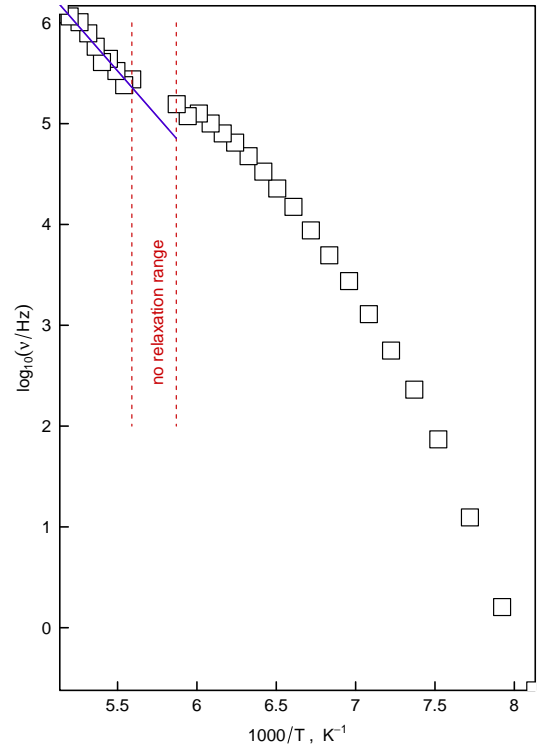}
\caption{Dielectric relaxation frequency vs. inverse temperature in ethanol at 0.5 GPa. ``No relaxation region'' corresponds to the temperature region, where solidification of supercooled ethanol takes place, and no relaxation in either glassy/supercooled liquid (lower temperatures) or solid phase (higher temperatures) is observed. Solid blue line is an Arrhenius extrapolation of relaxation frequency observed in the solid phase of ethanol to the lower temperatures.}
\label{f5}
\end{figure}

\begin{figure}
\includegraphics[width=\columnwidth]{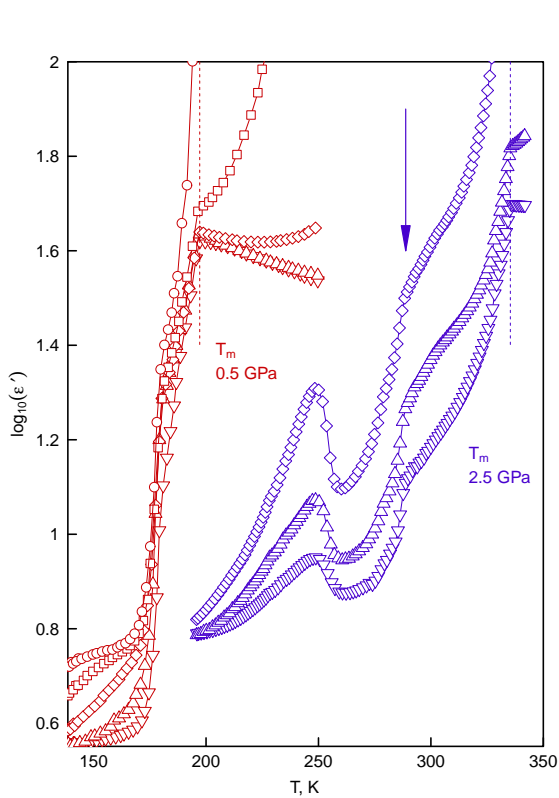}
\caption{Temperature dependence of ethanol dielectric permittivity in the low  (P=0.5 GPa, red curves) and high pressure (P=2.5 GPa, blue curves) phases. Various symbols correspond to different frequencies ($\circ$--100 Hz, $\square$-- 1 KHz,   $\lozenge$-- 10 kHz, $\triangledown$ -- 100 kHz, $\triangle$-- 1 MHz). Dashed vertical lines mark onset of melting temperature. Downward arrow marks transition from the high-pressure ordered phase to high-pressure orientationally disordered phase accompanied by the increase of dielectric permittivity.}
\label{f6}
\end{figure}

It is worth noticing, that our dielectric spectroscopy experiments suggest presence of slight orientational disorder in the vicinity of the melting temperature at ambient pressure (which resembles the onset of disorder in fully ordered $\alpha$-phase of methanol while approaching $\alpha$-$\beta$ phase transition). Still the overall behavior of dielectric relaxation is quite different from the methanol case int he premelting region and we can not designate distinct temperature region, in which ethanol transforms into disordered state. Since the ethanol easily vitrifies upon cooling at low pressures, it is easy to register both Debye and $\alpha$-relaxation processes, being in experimentally attainable frequency range.  However,  glassy state of ethanol is unstable and  crystallizes upon heating (the cold crystallization). Upon heating near the premelting region the disordering of ethanol crystal phase takes place, which manifests itself in gradual increase of dielectric permittivity in the low frequency region. The example of such a relaxation is shown in Fig.~\ref{f4}. This dielectric response in this temperature region is also frequency dependent and its extrapolated characteristic frequency slightly below the frequency of the Debye relaxation collected in the glassy/supercooled liquid state (see Fig.~\ref{f5}, where the data observed at elevated pressure $P=0.5$ GPa are presented). The characteristic frequency of relaxation process in glassy and solid phases of ethanol is obtained by simultaneous fitting of real and imaginary part of dielectric responce by simply ignoring polarization effects according to Equation~\ref{eps} with relaxation part described by Equation~\ref{cd1}. This quite matches the model proposed for description of relaxation in methanol but with very small downshift of characteristic frequency during solidification. Still we would stress once more, that the amplitude of this relaxation process monotonously depends on temperature and one can find no kinks on the temperature dependence of low-frequency dielectric permittivity which could be ascribed to phase transitions to disordered high-temperature phase (see Fig.~\ref{f6}). The amplitude of dielectric response gradually increases with temperature, approaching the melting transition.

However, with crossing the pressure where the phase transition into high-pressure crystal phase takes place, the dielectric response of solid ethanol drastically changes. We consider this change as the evidence, that high pressure phase of ethanol near premelting region is in the state of hindered orientational disorder similarly to $\beta$-phase of methanol. However the similarity between these two phases require further investigation which will be carried out in the next Section~\ref{et:hp}. 
\section{Dielectric spectroscopy of ethanol at high pressures}
\label{et:hp}
Drastic difference between the high pressure and low-pressure crystal phases of ethanol (obviously beside their different crystal structures \cite{allan:jsr01}) is that by cooling from the liquid we can not obtain glassy state in the pressure region of high pressure phase stability (above 2.1 GPa) \cite{kondrin:jcp14}. Stability region of high pressure phase of ethanol is marked as II in Fig.~\ref{f0} b). However, we will demonstrate, that beside that, there is also difference in dielectric response of high pressure crystal phase, which is not observed in the low pressure phase but makes the former similar to disordered methanol phase. 

\begin{figure}
\includegraphics[width=\columnwidth]{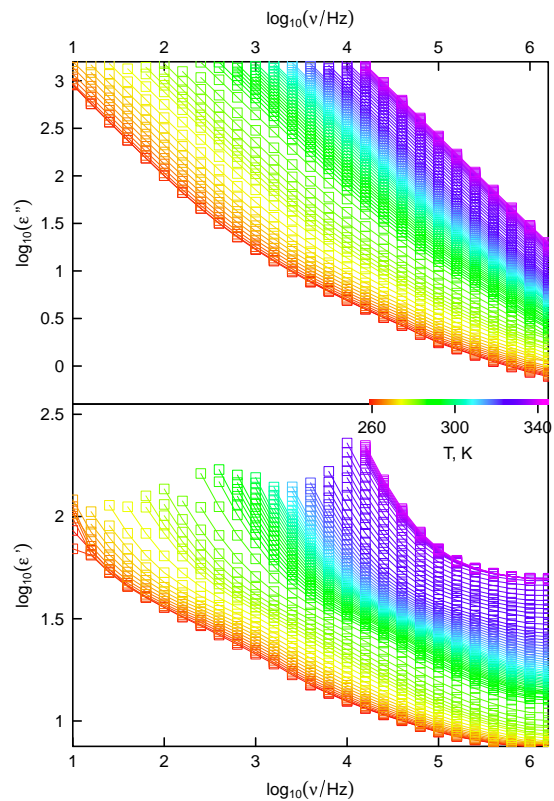}
\caption{Experimental dielectric relaxation in the liquid and solid high pressure phases of ethanol at 2.5 GPa.}
\label{f7}
\end{figure}

In Fig.~\ref{f7} typical example of dielectric response collected at pressure $P=2.5$ GPa is shown. In solid phase on the panel corresponding to real part of dielectric response the distinct step-like feature is evident. Amplitude of this feature is independent of temperature (while its characteristic frequency is increased with temperature rise). We can evaluate relative amplitude of this feature in respect to dielectric response of liquid ethanol as $(\varepsilon_L-\varepsilon_\infty)/(\varepsilon_S-\varepsilon_\infty)=(48.9-7.4)/(31.6-7.4)=1.7$. Still we observe, that the ratio is close to 2 (as in methanol) but substantially lower. In our opinion, this discrepancy can be connected to the significant hindrance of ethanol rotation in liquid phase, where longer hydrogen bonded threads might be present.  There is also a quite different behavior of dielectric permittivity at fixed frequencies in comparison to the data, collected at low pressures (see Fig.~\ref{f6} ). At these high-pressure slices a quite distinct kink (marked with arrow) is observed. which demarcates premelting region of high dielectric permittivity from  low temperature range, where low values of dielectric constant are observed. So in the premelting region, (which spans $T=288-335 $ K range)  dielectric response of solid ethanol resembles that of $\beta$-phase of solid methanol with constant (independent on temperature) value of static dielectric permittivity and high values of high-frequency dielectric constant $\varepsilon_\infty$. It is possible that these high-pressure phase of ethanol is similar to the orientationally disordered metastable form of ethanol with $fcc$ structure which is produced by special cooling treatment at ambient pressure \cite{srinivasan:prb96,ramos:prl97,jimenez:prb99,ramos:jpcm07}. In this metastable phase it was also found that the dielectric relaxation process is also  diminishes by two times at temperature lowering during transition to the orientational glass state  \cite{benkhof:jpcm98,jimenez:prb99}. This might indicate that the orientational disorder in this state also has hindered character.

We argue that the stability region of this partially orientationally high-pressure disordered phase is higher in temperature than the stability region of fully ordered phase II investigated by synchrotron radiation by Allan {\em et al.} \cite{allan:jsr01} at 2.75 GPa and room temperature. According to our high-frequency dielectric spectroscopy data the transition to the fully ordered phase takes place at temperatures 288 K at 2.5 GPa (marked by arrow in Fig.~\ref{f6}) and 318 K at 3 GPa \cite{kondrin:jcp14}. So it is highly likely that measurement of Allan {\em et al.} \cite{allan:jsr01} were carried at temperatures just below the transition into fully ordered phase of ethanol, so the lowering of pressure below 2.5 GPa at room temperature (but above 2.1 GPa) would result in transition into hindered plastic crystal phase of ethanol.

\section{Discussion}

So far we have demonstrated, that some sort of disorder present in  crystal phases of methanol and ethanol leads to significant increase of dielectric constant. Moreover, it was found, that in oprientationally disordered phase of methanol and high-pressure-high-temperature phase of ethanol it can lead to the values of dielectric constant as high as several tens of $\varepsilon$. The value of this constant practically does not depend on temperature, only characteristic frequency of relaxational process giving rize for such a large value of dielectric permittivity decreases with  lowering the temperature. However, common feature of all these phases is that the dielectric constant being approximately two times less than that of liquid in the premelting temperature region. We consider this an indication of hindered molecule rotation  in orientationally disordered phases in comparison to  liquid phase, and of the fact that in disordered crystal phase molecule's rotation is not free. We proposed to call such phases as hindered plastic crystals.

Further lowering of temperature induces  transition of these plastic crystal phases into fully ordered ones. So, the temperature stability region of hindered plastic crystal phases is not large and spans 10-20 K. In fully ordered phases at temperatures far from order-disorder  transition the dielectric constant  is small (below 10) which indicates  full extinguishing of relaxational process of molecules as a whole and is likely to be caused by induced polarization only. Beside low temperature phases of hindered plastic crystal the low pressure phase of ethanol is also fully ordered one. However, near the order-disorder transition we observe the increase of dielectric constant, which indicates thawing of rotational molecular degrees of freedom. So, fully ordered phases of lower-weight monoalcohols differ from hindered plastic crystal phases not only  in characteristic frequency of relaxational process, giving rize to large dielectric constant depends on temperature but its amplitude too.

It is interesting to compare dielectric response of hindered plastic crystals of ethanol and methanol to that of classical example of plastic crystal monoalcohol -- cyclooctanol (with entropy of fusion as low as 6.98 $J/(K~mole)$ \cite{sciesinski:pt95}). Its dielectric properties were investigated in number of works (see {\em e. g.} the works \cite{brand:prb97,michl:prl15,drozd-rzoska:m21,drozd-rzoska:jml22}). In contrast to high-pressure phase of ethanol and $\beta$-phase of methanol , dielectric constant of cyclooctanol does not decrease, but slightly increases during freezing of   liquid phase \cite{drozd-rzoska:m21}. This is accompanied with slight decrease (less than one order) of characteristic frequency of relaxational process \cite{drozd-rzoska:jml22}. So, this indicates that in contrast to methanol and ethanol, rotation of cyclooctanol molecule in plastic crystal phase is practically free. In this respect it resembles ordinary hexagonal ice. However, in our opinion, the cause of absence of hinderance of molecular rotations in cyclooctanol is different and may be due to the presence of large molecular ``tail'' of cyclooctanol. Interaction of these tails prevents formation of more tightly bound hydrogen-bonded chains in plastic crystal phase of cyclooctanol. So, in crystal phase of cyclooctanol there is no definite direction, along which the  hydrogen bonds can align.

There is another similarity between water and cyclooctanol in regard of their transition into fully ordred phase.  Hexagonal ice is notorious for difficulty of proton ordering  at low temperatures,  and it usually requires addition of external components (like ionic salt LiCl) to facilitate it. Similar situation is true for cyclooctanol, where orientational disorder tends to ``vitrify'' in process of lowering  the temperature . It is definetely opposite to ethanol and methanol, where hindered plastic crystal phases exist in very limited temperature range. How it can be connected with the hindrance of molecular rotation, is the question  yet to be resolved.
\section{Conclusions and final remarks}

Our first main conclusion is that static dielectric permittivity of methanol in orientationally disordered $\beta$-phase is significantly higher than supposed before. Our measurements in the premelting region suggest, that in disordered solid  phase of methanol, relaxational process typical to reorientation motion of molecules still exists, but significantly shifts to the low frequency region. Due to the hindered motion of molecules in this solid phase in comparison to almost free motion  in  liquid phase, the amplitude of dielectric relaxation in solid $\beta$-phase of ethanol is almost twice lower, than relaxation amplitude in liquid phase near melting temperature. This is different from ordinary ambient-pressure water, where dielectric relaxation shift to the low-frequency region upon crystallization is not accompanied with lowering of amplitude of this relaxation. This enables us to distinguish between true plastic crystals (like ice), where rotation of the molecules' dipole moment is free in crystal phase and hindered plastic crystal (like methanol), where rotation of dipole moment is restricted to certain plane. In methanol, the axis normal to this plane is the line, connecting carbon and oxygen atoms.

We argue, that the phase similar to hindered plastic crystal phase of methanol exists at high pressures in another small-molecule monoalcohol -  ethanol too. Low pressure phase of ethanol is obviously ordered, but application of high pressure (above 2.1 GPa) and temperatures leads to orientational disordering of its molecules. We have found, that amplitude ratio of relaxational process observed in solid high-pressure high-temperature phase of ethanol relative to one observed in liquid phase is close to two, but substantially lower (approximately equal to 1.7). We believe, that the decrease of this coefficient  indicates somehow hindered motion of ethanol molecules in high-pressure liquid phase,  compared to methanol.

We should also mention that these results were only possible due to our experimental setup which allows us to collect dielectric responce during simultaneous heating and scanning the frequency. Also there was significant breakthrough in data processing procedure that enables us to separate electrode polarization effects from the dielectric relaxation of the sample in the low frequency region and unambiguously establish the value of static dielectric constants in the solid phases of ethanol and methanol.

\input{hind-pc6.bbl}
\end{document}

%% file: hind-pc6.bbl
%